\documentclass[prl,aps,twocolumn,preprintnumbers,amssymb,nobibnotes,nofootinbib,superscriptaddress,raggedbottom,10pt]{revtex4-1}
\usepackage{hyperref,graphicx,array,multirow,color,amsmath,mathrsfs,titlesec,shuffle,tikz}
\usetikzlibrary{calc,decorations.markings}
\titleformat{\section}{\centering\normalsize\normalfont\bf}{\thesection}{0em}{}
\hypersetup{pdftitle={},pdfcreator={},linkcolor=[rgb]{0.15,0.35,0.75},colorlinks=true,citecolor=[rgb]{0.675,0,0.2},urlcolor=[rgb]{0.15,0.35,0.65}}
\DeclareMathOperator*{\Res}{\mathrm{Res}}
\newcommand{\fwbox}[2]{\text{\makebox[#1][c]{$\hspace{-150pt}\displaystyle#2\hspace{-150pt}$}}}
\newcommand{\fwboxL}[2]{\text{\makebox[#1][l]{$#2$}}}
\newcommand{\fwboxR}[2]{\text{\makebox[#1][r]{$#2$}}}

\newcommand{\eq}[1]{\vspace{-3.5pt}\begin{equation}\hspace{2pt}#1\hspace{-0pt}\vspace{-3.5pt}\end{equation}}
\newcommand{\eqL}[1]{\eq{\fwboxL{0pt}{\hspace{-115pt}#1}}}
\newcommand{\fig}[3]{\raisebox{#1}{\includegraphics[scale=#2]{#3}}}

\newcommand{\ab}[1]{\langle #1\rangle}
\newcommand{\equivR}{\fwbox{13.5pt}{\hspace{-0pt}\fwboxR{0pt}{\raisebox{0.47pt}{\hspace{1.45pt}:\hspace{-3pt}}}=\fwboxL{0pt}{}}}
\newcommand{\equivL}{\fwbox{13.5pt}{\fwboxR{0pt}{}=\fwboxL{0pt}{\raisebox{0.47pt}{\hspace{-3pt}:\hspace{1.45pt}}}}}

\newcommand{\newcap}{\mathrm{\raisebox{0.75pt}{{$\,\bigcap\,$}}}}

\newcommand{\tncap}{\scalebox{0.8}{$\!\newcap\!$}}

\definecolor{hblue}{rgb}{0,0,0.575}
\definecolor{hred}{rgb}{0.575,0.0,0.225}
\definecolor{hteal}{rgb}{0.0,0.545,0.7451}
\definecolor{perm}{rgb}{0.1,0.45,0.85}

\renewcommand{\r}[1]{{\color{hred}#1}}
\renewcommand{\b}[1]{{\color{hblue}#1}}
\renewcommand{\phi}{\varphi}
\renewcommand{\hat}{\widehat}

\begin{document}
\title{\texorpdfstring{An Elliptic, Yangian-Invariant `Leading Singularity'\\[-12pt]~}{An Elliptic Yangian-Invariant, `Leading Singularity'}}
\author{Jacob~L.~Bourjaily}%\email{bourjaily@psu.edu}
\affiliation{Institute for Gravitation and the Cosmos, Department of Physics,\\Pennsylvania State University, University Park, PA 16802, USA}
\affiliation{Niels Bohr International Academy and Discovery Center, Niels Bohr Institute,\\University of Copenhagen, Blegdamsvej 17, DK-2100, Copenhagen \O, Denmark}
\author{Nikhil~Kalyanapuram}%\email{nkalyanapuram@psu.edu}
\affiliation{Institute for Gravitation and the Cosmos, Department of Physics,\\Pennsylvania State University, University Park, PA 16802, USA}
\author{Cameron~Langer}%\email{ckl5552@psu.edu}
\affiliation{Institute for Gravitation and the Cosmos, Department of Physics,\\Pennsylvania State University, University Park, PA 16802, USA}
\author{Kokkimidis~Patatoukos}%\email{kzp326@psu.edu}
\affiliation{Institute for Gravitation and the Cosmos, Department of Physics,\\Pennsylvania State University, University Park, PA 16802, USA}
\author{Marcus~Spradlin}%\email{marcus\_spradlin@brown.edu}
\affiliation{Department of Physics, Brown University, Providence, RI 02912, USA}
\affiliation{Brown Theoretical Physics Center, Brown University, Providence, RI 02912, USA}

%%%%%%%%%%%%%%%%%%%%%%%%%%%%%%%%%%%%%%%%%%%%%%%%%%%%%%%%%%%%%%%%%%%%%%%%%%%%%%%%%%%%%%%%%%%

%%%%%%%%%%%%%%%%%%%%%%%%%%%%%%%%%%%%%%%%%%%%%%%%%%%%%%%%%%%%%%%%%%%%%%%%%%%%%%%%%%%%%%%%%%%
\begin{abstract}
We derive closed formulae for the first examples of non-algebraic, elliptic `leading singularities' in planar, maximally supersymmetric Yang-Mills theory and show that they are Yangian-invariant.%
\end{abstract}
\maketitle
%%%%%%%%%%%%%%%%%%%%%%%%%%%%%%%%%%%%%%%%%%%%%%%%%%%%%%%%%%%%%%%%%%%%%%%%%%%%%%%%%%%%%%%%%%%

%%%%%%%%%%%%%%%%%%%%%%%%%%%%%%%%%%%%%%%%%%%%%%%%%%%%%%%%%%%%%%%%%%%%%%%%%%%%%%%%%%%%%%%%%%%
\vspace{-15pt}\section{Introduction}\label{introduction_section}\vspace{-14pt}
%%%%%%%%%%%%%%%%%%%%%%%%%%%%%%%%%%%%%%%%%%%%%%%%%%%%%%%%%%%%%%%%%%%%%%%%%%%%%%%%%%%%%%%%%%%
%
The \emph{residues} of (the loop integrands of) scattering amplitudes on iterated, simple poles have played a critical and starring role through most of the incredible advances in our understanding of perturbative quantum field theory in recent years. Such residues, when they are of maximal co-dimension, have been called `leading singularities' \cite{ELOP,Cachazo:2008vp} and have been used in the context of generalized unitarity \cite{Bern:1994zx,Bern:1994cg,Bern:1995db,Britto:2004nc} to construct integrands for amplitudes to impressive orders of perturbation \cite{Britto:2004nc,Buchbinder:2005wp,Bern:2007hb,Bern:2007ct,Cachazo:2008dx,Cachazo:2008hp,Spradlin:2008uu,Bourjaily:2011hi,Bourjaily:2013mma,Bourjaily:2015bpz,Bourjaily:2015jna,Bourjaily:2016evz,Bourjaily:2017wjl,Bourjaily:2018omh,Bourjaily:2019gqu,Bourjaily:2019iqr}. Leading singularities have also appeared as the individual terms generated by BCFW recursion relations for amplitudes at tree-level \cite{Britto:2004ap,Britto:2005fq} and beyond \cite{CaronHuot:2010zt,ArkaniHamed:2010kv,Boels:2011tp}. 

In the case of maximally supersymmetric Yang-Mills theory in the planar limit (sYM), leading singularities were discovered to enjoy an infinite-dimensional Yangian symmetry algebra \cite{Drummond:2008vq,Alday:2008yw,Drummond:2009fd,Drummond:2010qh}, and were later found to be classified according to the `positroid' stratification of Grassmannian manifolds \cite{Postnikov:2006,Williams:2003a,ArkaniHamed:book}. In \cite{Drummond:2010uq}, the correspondence between Yangian-invariants and compact contour integrals over the positroid-canonical top-form \cite{ArkaniHamed:2009dn,ArkaniHamed:2009vw,Mason:2009qx} in the Grassmannian was established. 

As maximal co-dimension residues, leading singularities can be viewed as the coefficients of loop amplitude integrands that are locally (in loop-momentum space) `$d\!\log$' differential forms---wedge-products of simple poles in all loop-momentum variables. (A multidimensional residue can in fact be defined by this fact alone---given by the inverse Jacobian of the requisite change of variables, evaluated at this point.) It is no surprise, therefore, that leading singularities also appear as the coefficients of polylogarithmic functions arising from `loop integration' (over the Feynman-contour of real momenta). Indeed, the connection between the \emph{sufficiency} of leading singularities as information required to construct amplitudes and the existence of `$d\!\log$' representations of loop integrands has been the subject of much research in recent years \cite{ArkaniHamed:2010gh,Bourjaily:2013mma,Arkani-Hamed:2014via,Bourjaily:2017wjl}. 

Nevertheless, an increasing body of research suggests that many perturbative scattering amplitudes are not in fact polylogarithmic \cite{Broadhurst:1993mw,Bloch:2013tra,CaronHuot:2012ab,Bourjaily:2017bsb}, and can involve a much richer variety of functions---such as elliptic multiple polylogarithms \cite{Remiddi:2017har,Broedel:2017kkb,Broedel:2017siw}, or even integrals over Calabi-Yau manifolds \cite{Brown:2010bw,Bourjaily:2018ycu,Bourjaily:2018yfy,Bourjaily:2019hmc}. When the integrands of loop amplitudes are not $d\!\log$ somewhere locally, `leading singularities' as they have been so-far defined represent insufficient information (for reconstruction via unitarity) about perturbative amplitudes. The attempts to use generalized unitarity in such cases have relied on less intrinsically well-defined (or well-motivated) strategies---such as matching amplitude integrands functionally through some number of off-shell evaluations (see e.g.\ \cite{Bourjaily:2015jna,Bourjaily:2017wjl,Bourjaily:2020qca}).\\[-10pt]

In this Letter, we argue that the notion of `leading singularity' must be broadened to include \emph{any} full-dimensional compact, contour integral of an amplitude. Up to factors of $(2\pi i)$, this definition automatically includes those functions covered previously,  but expands their scope to include contours that are not merely the products of circles enclosing simple poles. We conjecture that this new, broadened definition of `leading singularities' represents \emph{complete} unitarity-level information for amplitudes in perturbative sYM (and perhaps considerably beyond).\\[-10pt]

The first appearance of non-polylogarithmic structure for planar sYM theory occurs in the 10-particle N$^3$MHV amplitude at two loops \cite{CaronHuot:2012ab}. In this Letter, we derive closed formulae for the elliptic leading singularities of this amplitude. We have checked that they are Yangian-invariant, and very non-trivially, since they involve the periods of complete elliptic integrals. Thus, they are not any elliptic analogues of `residues'---which would be necessarily algebraic---but rather contour integrals directly, carrying (some notion of) transcendental weight. Interestingly, the loop integrand whose coefficient would be an elliptic leading singularity takes the form of what has been defined as a `pure' elliptic integral \cite{Broedel:2018qkq}. We strongly suspect that this magic is not accidental: namely, that for any basis of loop integrands diagonalized on maximal-dimension, compact contours, the coefficients of amplitudes (in sYM) will be Yangian invariant and all integrands will be `pure' (in a sense suitably generalized to integrals involving higher-dimensional Calabi-Yau periods). But we leave such speculations to future work.

%%%%%%%%%%%%%%%%%%%%%%%%%%%%%%%%%%%%%%%%%%%%%%%%%%%%%%%%%%%%%%%%%%%%%%%%%%%%%%%%%%%%%%%%%%%
\vspace{-12pt}\section{Elliptic \emph{Sub}-Leading Singularities of sYM}\vspace{-14pt}
%%%%%%%%%%%%%%%%%%%%%%%%%%%%%%%%%%%%%%%%%%%%%%%%%%%%%%%%%%%%%%%%%%%%%%%%%%%%%%%%%%%%%%%%%%%
%
We begin our analysis with the `double-box' cut of the 2-loop, 10 particle N$^3$MHV amplitude in planar sYM: 
\eq{\fwboxR{0pt}{\mathfrak{db}_{\pm}(\r{\alpha})\equivR}\fig{-28pt}{1}{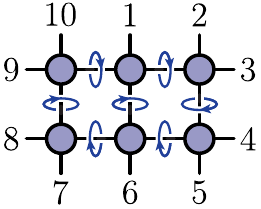}\label{double_box_heptacut}}
It is a `next-to-leading singularity' in the ordinary sense (a co-dimension 7 residue), and corresponds to a contour which encircles the seven poles from the shown propagators. This residue depends on one internal (on-shell) degree of freedom denoted $\r{\alpha}$ and a discrete label `$\pm$' signifying which of the two branches of the solution to the cut equations is taken. (As a positroid configuration, it corresponds to a 13-dimensional cell in $G_+(3,10)$ labeled by the decorated permutation ${\color{perm}\{6,5,4,8,7,11,10,9,13,12\}}$, integrated against the 12 constraints $\delta^{3\!\times\!4}\!\big(C\!\cdot\!Z\big)$---leaving a one-form over the remaining variable (see e.g.\ \cite{ArkaniHamed:book,Bourjaily:2012gy}).)

We can give an explicit formula for the double-box cut $\mathfrak{db}_{\pm}(\r{\alpha})$ as a single residue of (the relevant term of) the co-dimension 6 `kissing-triangle' function given in \cite{Bourjaily:2015jna}:
\eqL{d{\color{hred}\alpha}\,d{\color{hred}\beta}\frac{1}{{\color{hred}\alpha}\,{\color{hred}\beta}}R\big[\b{1},{\color{hred}\hat{2}},{\color{hred}\hat{6}},{\color{hred}\hat{7}},{\color{hred}\hat{1}}\big]R\big[{\color{hred}\hat{2}},{\color{hblue}3},{\color{hblue}4},{\color{hblue}5},{\color{hblue}6}\big]R\big[{\color{hred}\hat{7}},{\color{hblue}8},{\color{hblue}9},{\color{hblue}10},{\color{hblue}1}\big]\label{kissing_triangle_formula}}
where $R[\b{a},\!\b{b},\!\b{c},\!\b{d},\!\b{e}]$ is the $R$-invariant \cite{Drummond:2008cr,Bullimore:2010pj} with momentum supertwistor arguments $\{\b{\mathcal{Z}_a},\ldots,\b{\mathcal{Z}_e}\}$  \cite{Hodges:2009hk}, and the `hatted'  arguments are defined geometrically (see e.g.\ \cite{ArkaniHamed:2010gh}) by:
\eqL{\begin{array}{l@{$\hspace{20pt}$}l}{\color{hred}\hat{2}}({\color{hred}\alpha})\,\equivR{\color{hblue}\mathcal{Z}_2}+{\color{hred}\alpha}{\color{hblue}\mathcal{Z}_1}&{\color{hred}\hat{7}}({\color{hred}\beta})\,\equivR{\color{hblue}\mathcal{Z}_7}+{\color{hred}\beta}\,{\color{hblue}\mathcal{Z}_6}\\
{\color{hred}\hat{6}}({\color{hred}\alpha})\,\equivR\big({\color{hblue}5}\,{\color{hblue}6}\big)\tncap\big({\color{hred}\hat{2}}\,{\color{hblue}3}\,{\color{hblue}4}\big)&{\color{hred}\hat{1}}({\color{hred}\beta})\,\equivR\big({\color{hblue}10}\,{\color{hblue}1}\big)\tncap\big({\color{hred}\hat{7}}\,{\color{hblue}8}\,{\color{hblue}9}\big)\,.\end{array}}
(Notice that the variables $\r{\alpha}$ and $\r{\beta}$ both carry non-trivial little-group weights. These could be absorbed by rescaling them (and the form) appropriately; but any such rescaling would cancel out of any complete compact contour integral (or residue).)

To obtain (\ref{double_box_heptacut}) from (\ref{kissing_triangle_formula}), we must compute the residue of (\ref{kissing_triangle_formula}) associated with the contour encircling the pole $\ab{\r{\hat{2}\,\hat{6}\,\hat{7}\,\hat{1}}}=0$. We may take this residue with respect to either variable, and we choose to eliminate $\r{\beta}$. And because $\ab{\r{\hat{2}\,\hat{6}\,\hat{7}\,\hat{1}}}$ is quadratic in $\r{\beta}$, there are two branches available. 

For the sake of concreteness, we note that 
\eq{\Res_{\ab{\r{\hat{2}\,\hat{6}\,\hat{7}\,\hat{1}}}=0}\left[\frac{d\,\r{\beta}}{\ab{\r{\hat{2}\,\hat{6}\,\hat{7}\,\hat{1}}}}\right]=\pm\frac{c_y}{\,y(\r{\alpha})}\,}
where $y(\r{\alpha})^2$ is the (by-construction \emph{monic}) quartic polynomial
\eqL{\begin{split}\hspace{-10pt}\frac{1}{c_y^2}\,y^2(\r{\alpha})\equivR\!&\Big(\!\ab{{\color{hred}\hat{2}}\,{\color{hred}\hat{6}}\,{\color{hblue}6}\,\big({\color{hblue}7}\,{\color{hblue}8}\,{\color{hblue}9}\big)\tncap\big({\color{hblue}10}\,{\color{hblue}1}\big)}{+}\ab{{\color{hred}\hat{2}}\,{\color{hred}\hat{6}}\,{\color{hblue}7}\,\big({\color{hblue}6}\,{\color{hblue}8}\,{\color{hblue}9}\big)\tncap\big({\color{hblue}10}\,{\color{hblue}1}\big)}\!\Big)^{\!2}\\[-5pt]
&{-}4\,\ab{{\color{hred}\hat{2}}\,{\color{hred}\hat{6}}\,{\color{hblue}6}\,\big({\color{hblue}6}\,{\color{hblue}8}\,{\color{hblue}9}\big)\tncap\big({\color{hblue}10}\,{\color{hblue}1}\big)}\ab{{\color{hred}\hat{2}}\,{\color{hred}\hat{6}}\,{\color{hblue}7}\,\big({\color{hblue}7}\,{\color{hblue}8}\,{\color{hblue}9}\big)\tncap\big({\color{hblue}10}\,{\color{hblue}1}\big)} \nonumber%\label{defn_of_y2}
\end{split}}
where $c_y^2$ is defined to be the inverse of the coefficient of $\r{\alpha}^4$ on the right hand side of the definition above (so as to render $y^2(\r{\alpha})$ monic), and the particular solutions $\r{\beta}^*_{\pm}$ to $\ab{\r{\hat{2}\,\hat{6}\,\hat{7}\,\hat{1}}}=0$ are given by 
\eqL{\hspace{-7.5pt}\r{\beta}^*_{\pm}\!\equivR\!\!\!\frac{\ab{{\color{hred}\hat{2}}\,{\color{hred}\hat{6}}\,{\color{hblue}6}\,\big({\color{hblue}7}\,{\color{hblue}8}\,{\color{hblue}9}\big)\tncap\big({\color{hblue}10}\,{\color{hblue}1}\big)}{\!+\!}\ab{{\color{hred}\hat{2}}\,{\color{hred}\hat{6}}\,{\color{hblue}7}\,\big({\color{hblue}6}\,{\color{hblue}8}\,{\color{hblue}9}\big)\tncap\big({\color{hblue}10}\,{\color{hblue}1}\big)}\!\!\pm\! y(\hspace{-0.5pt}\r{\alpha}\hspace{-0.5pt})/c_y}{2\ab{{\color{hred}\hat{2}}\,{\color{hred}\hat{6}}\,\big({\color{hblue}6}\,{\color{hblue}8}\,{\color{hblue}9}\big)\tncap\big({\color{hblue}10}\,{\color{hblue}1}\big)\,{\color{hblue}6}}}\!.\nonumber}
To define the double-box residue (\ref{double_box_heptacut}), therefore, we may replace the pole $\ab{\r{\hat{2}\,\hat{6}\,\hat{7}\,\hat{1}}}$ (appearing in the denominator of $R[\b{1},\r{\hat{2}},\r{\hat{6}},\r{\hat{7}},\r{\hat{1}}]$) with $y(\r{\alpha})$, and replace $\r{\beta}$ with $\r{\beta}^*_{\pm}(\r{\alpha})$ everywhere else in (\ref{kissing_triangle_formula}). 

As $y^2(\r{\alpha})$ is an irreducible quartic, the differential forms $\mathfrak{db}_{\pm}(\r{\alpha})$ should be understood as involving the geometry of an elliptic curve. In general, any such differential form may be represented in the form
\eq{\mathfrak{db}_{\pm}(\r{\alpha})\equivL d\r{\alpha}\Big[Q(\r{\alpha})\pm\frac{1}{y(\r{\alpha})}R(\r{\alpha})\Big]}
where $Q$ and $R$ are algebraic (super) functions of $\r{\alpha}$. As we are interested in taking a contour integral over the cycles of the elliptic curve, only the term involving $1/y(\r{\alpha})$ matters; we may extract this piece by writing\\[-10pt]
\eq{\Big[\mathfrak{db}_{+}(\r{\alpha})-\mathfrak{db}_{-}(\r{\alpha})\Big]\!\equivL\mathfrak{db}(\r\alpha)\equivL\frac{d\r\alpha}{y(\r\alpha)}\hat{\mathfrak{db}}(\r{\alpha})\,.}
% 

%%%%%%%%%%%%%%%%%%%%%%%%%%%%%%%%%%%%%%%%%%%%%%%%%%%%%%%%%%%%%%%%%%%%%%%%%%%%%%%%%%%%%%%%%%%
\vspace{-16pt}\subsection{Analytic Structure of the Elliptic Form $\mathfrak{db}(\r{\alpha})$}\vspace{-14pt}
%%%%%%%%%%%%%%%%%%%%%%%%%%%%%%%%%%%%%%%%%%%%%%%%%%%%%%%%%%%%%%%%%%%%%%%%%%%%%%%%%%%%%%%%%%%
%
It is not hard to see that the differential form $\mathfrak{db}(\r{\alpha})$ has many simple poles---corresponding to the various factorization channels of the 6 four-particle amplitudes appearing in (\ref{double_box_heptacut}). Every such factorization channel has the topology of a `pentabox' leading singularity; counting every distinct solution to the cut equations for each topology, there are 24 total such `boundary' on-shell functions; let us denote them $\mathfrak{pb}_i$. 

Each of these `factorizations' of the double-box cut (\ref{double_box_heptacut}) corresponds to a simple pole located at $a_i$ in the $\r{\alpha}$-plane with residue \emph{equal} to the corresponding pentabox on-shell function $\mathfrak{pb}_i$. For example, near
\eq{\r{\alpha}\to a_1\equivR\frac{\ab{{\color{hblue}2}\,({\color{hblue}3}\,{\color{hblue}4})\tncap({\color{hblue}10}\,{\color{hblue}1}\,{\color{hblue}2})\,{\color{hblue}5}\,{\color{hblue}6}}}{\ab{({\color{hblue}3}\,{\color{hblue}4})\tncap({\color{hblue}10}\,{\color{hblue}1}\,{\color{hblue}2})\,{\color{hblue}5}\,{\color{hblue}6}\,{\color{hblue}1}}}\,,}
the differential form $\mathfrak{db}(\r{\alpha})$ has a simple pole with residue\\[-10pt]
\eq{\fwboxR{0pt}{\Res_{\r{\alpha}=a_1}\Big[\mathfrak{db}(\r{\alpha})\Big]=}\fig{-32pt}{1}{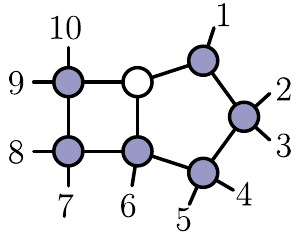}\fwboxL{0pt}{\equivL\mathfrak{pb}_1\,.}}
This function corresponds to one of the ordinary co-dimension one boundaries of the positroid configuration of the double-box; as such it can easily be computed as the canonical (12-dimensional) form in the Grassmannian integrated against $\delta^{3\times4}\!\big(C\!\cdot\!\b{Z})$. Regardless of how it is represented or computed, the location of each pole and its residue is easy to determine explicitly. (Interestingly, it is worth noting that all of the poles $a_i$ lie on the real axis for positive kinematics (and are all negative semi-definite).) These have been given explicitly in the ancillary files to this work. 

Expanding $\mathfrak{db}(\r{\alpha})$ into a basis of forms with manifest simple poles results in a representation of $\mathfrak{db}(\r{\alpha})$ which may be written
\eq{\mathfrak{db}(\r{\alpha})\equivL\frac{d\r{\alpha}}{y(\r{\alpha})}\mathfrak{db}_0+d{\r{\alpha}}\sum_{i=1}^{24}\frac{y(a_i)}{(\r{\alpha}{-}a_i)y(\r{\alpha})}\mathfrak{pb}_i\label{analytic_formula_for_db_form}}
where $\mathfrak{db}_0$---the coefficient of the differential form $d\r{\alpha}/y(\r{\alpha})$ \emph{without} any simple poles---is therefore defined indirectly (but explicitly, and without ambiguity) by 
\eq{\mathfrak{db}_0\equivR \hat{\mathfrak{db}}(\r{\alpha})-\sum_{i=1}^{24}\frac{y(a_i)}{(\r{\alpha}{-}a_i)}\mathfrak{pb}_i.\label{first_pass_dbhat}}
(Actually, for (\ref{analytic_formula_for_db_form}) and (\ref{first_pass_dbhat}), there is in fact (exactly) one pole at $\r{\alpha}\!=\!-\infty$; for this term, the differential form in the sum should be understood as being $d\r{\alpha}\,\frac{\r{\alpha}}{y(\r{\alpha})}$.) 

Importantly, since $\mathfrak{db}_0$ has no poles in $\r{\alpha}$ (including at infinity), it must be \emph{independent} of $\r{\alpha}$!---a fact that we have checked analytically. As such, it is worthwhile to express it in the form
\eq{\mathfrak{db}_0\equivR \hat{\mathfrak{db}}(\r{\alpha^*})-\sum_{i=1}^{24}\frac{y(a_i)}{(\r{\alpha^*}{-}a_i)}\mathfrak{pb}_i\label{bare_piece_final_form}}
for any choice of $\r{\alpha^*}$. As every expression appearing in the right-hand side of (\ref{bare_piece_final_form}) is fully-known analytically (as superfunctions (or expressed in terms of $R$-invariants)), this provides a concrete definition for $\mathfrak{db}_0$.

%%%%%%%%%%%%%%%%%%%%%%%%%%%%%%%%%%%%%%%%%%%%%%%%%%%%%%%%%%%%%%%%%%%%%%%%%%%%%%%%%%%%%%%%%%%
\vspace{-12pt}\section{Elliptic Leading Singularities of sYM}\vspace{-14pt}
%%%%%%%%%%%%%%%%%%%%%%%%%%%%%%%%%%%%%%%%%%%%%%%%%%%%%%%%%%%%%%%%%%%%%%%%%%%%%%%%%%%%%%%%%%%
%
We are now in a position to determine the elliptic `leading singularity'---the integral of the form $\mathfrak{db}(\r{\alpha})$ over some choice of elliptic cycle, say $\Omega_a$:
\eqL{\mathfrak{e}_a\equivR\int_{\Omega_a}\!\!\mathfrak{db}(\r{\alpha})=\int_{\Omega_a}\!\!\frac{d\r{\alpha}}{y(\r\alpha)}\,\hat{\mathfrak{db}}(\r{\alpha})=\pm2\int_{\Omega_a}\!\!\mathfrak{db}_{\pm}(\r\alpha)\,.\label{elliptic_ls_v0}}
To specify the particular elliptic cycle $\Omega_a$, it is worthwhile that to note that for positive (non-degenerate) momentum-twistor kinematics (see e.g.\ \cite{Arkani-Hamed:2013jha} for a discussion of positive kinematics), it turns out that the roots $r_i$ of the quartic $y^2(\r{\alpha})$,
\eqL{y^2(\r\alpha)\equivL(\r\alpha-r_1)(\r\alpha-r_2)(\r\alpha-r_3)(\r\alpha-r_4)\,,}
always come in complex conjugate pairs---between which we may introduce branch cuts. To be clear, we have chosen to order the roots such that $r_1^*\!\equivL r_2$ and $r_3^*\!\equivL r_4$, with $\mathfrak{Re}(r_1)\!>\!\mathfrak{Re}(r_3)$ and $\mathfrak{Im}(r_{1,3})\!>\!0$; with this ordering of the roots (the reverse of the default ordering from `\texttt{Solve[]}' in \textsc{Mathematica}), the cross-ratio 
\eq{\phi\equivR\frac{(r_2{-}r_1)(r_3{-}r_4)}{(r_2{-}r_3)(r_1{-}r_4)}}
is always real; moreover $\phi\in[0,1]$ for positive $\b{\mathcal{Z}}$s. With these conventions, we define $\Omega_a$ to be the contour `enclosing' the branch cut between the complex-conjugate pair of roots $r_{1,2}$ which does \emph{not} encircle any of the simple poles of $\mathfrak{db}(\r\alpha)$.

In order to compute the elliptic leading singularity (\ref{elliptic_ls_v0}) therefore, we merely need to note the basic period integrals appearing in (\ref{analytic_formula_for_db_form})
\eq{\int_{\Omega_a}\!\!\!d\r\alpha\,\,\frac{1}{y(\r\alpha)}=\frac{4\,i}{\sqrt{(r_3{-}r_2)(r_4{-}r_1)}}K[\phi]\label{first_kind_period_a}}
and 
\begin{align}\int_{\Omega_a}\!\!\!d\r\alpha\,\,\frac{y(a_i)}{(\r\alpha{-}a_i)y(\r\alpha)}=&\frac{4\,i}{\sqrt{(r_3{-}r_2)(r_4{-}r_1)}}\Bigg(\label{third_kind_period_a}\\&\hspace{-100pt}\frac{y(a_i)}{(r_4{-}a_i)}K[\phi]+\frac{y(a_i)(r_4{-}r_2)}{(r_2{-}a_i)(r_4{-}a_i)}\Pi\left[\frac{(r_4{-}a_i)(r_2{-}r_1)}{(r_2{-}a_i)(r_4{-}r_1)},\phi\right]\!\!\Bigg)\nonumber\end{align}
where $K[\phi]$ and $\Pi[q,\phi]$ are the complete elliptic integrals of the first and third kinds, respectively, defined in accordance with the conventions of \textsc{Mathematica}. Of particular importance is that both (\ref{first_kind_period_a}) and (\ref{third_kind_period_a}) are pure-imaginary for positive kinematics; for the latter integral (\ref{third_kind_period_a}), this statement is nontrivial (even for $\phi\in\in[0,1]$) as the coefficients of both complete elliptic integrals appearing in (\ref{third_kind_period_a}) have non-zero real and imaginary parts, and only the combination is pure-imaginary. The full integral in (\ref{elliptic_ls_v0}) is obtained by integrating each term in (\ref{analytic_formula_for_db_form}), and using the explicit formula for $\mathfrak{db}_0$ in (\ref{bare_piece_final_form}) (for any choice of $\r\alpha^*$). As a consequence of the above discussion, this representation of $\mathfrak{e}_a$ is term-by-term purely imaginary.

One reason for our preference for the $a$-cycle (and also for our conventions regarding the roots) is that in the space of positive kinematics, the only possible kinematic degenerations at co-dimension one result in the collision of one of the two pairs of complex-conjugate roots. When this happens, it is easy to see that \emph{both} integrals (\ref{first_kind_period_a}) and (\ref{third_kind_period_a}) become \emph{equal to} $(2\pi i)$ times the \emph{residue} around the corresponding simple pole generated by the colliding pair of roots. (Recall that $K[0]=\Pi[0,0]=\pi/2$.) The $b$-cycle integrals, in contrast, diverge upon such degenerations.

%%%%%%%%%%%%%%%%%%%%%%%%%%%%%%%%%%%%%%%%%%%%%%%%%%%%%%%%%%%%%%%%%%%%%%%%%%%%%%%%%%%%%%%%%%%
\vspace{-16pt}\subsection{More Concise Formulae for the Leading Singularities}\vspace{-14pt}
%%%%%%%%%%%%%%%%%%%%%%%%%%%%%%%%%%%%%%%%%%%%%%%%%%%%%%%%%%%%%%%%%%%%%%%%%%%%%%%%%%%%%%%%%%%
%
In the discussion above, the reader should notice that every pentabox on-shell function $\mathfrak{pb}_i$ appears twice: once in the definition of $\mathfrak{db}_0$ in (\ref{bare_piece_final_form}) (where $\r{\alpha^*}$ may be taken as arbitrary) and once as the coefficient of the a particular ($\r\alpha$-dependent) differential form in $\mathfrak{db}(\r\alpha)$ in (\ref{analytic_formula_for_db_form}). From the first, (\ref{first_kind_period_a}) results in a contribution to $\mathfrak{e}_a$ of 
\eqL{\mathfrak{e}_a\supset\b{-}\frac{4\,i}{\sqrt{(r_3{-}r_2)(r_4{-}r_1)}}K[\phi]\times\frac{y(a_i)}{(\r{\alpha^*}{-}a_i)}\mathfrak{pb}_i\,;}
and from the second, (\ref{third_kind_period_a}) results in a contribution of\\[-14pt]
\begin{align}\mathfrak{e}_a\supset&\frac{4\,i}{\sqrt{(r_3{-}r_2)(r_4{-}r_1)}}\Bigg(\frac{y(a_i)}{(\r{r_4{-}}a_i)}K[\phi]\\&+\frac{y(a_i)(r_4{-}r_2)}{(r_2{-}a_i)(r_4{-}a_i)}\Pi\left[\frac{(r_4{-}a_i)(r_2{-}r_1)}{(r_2{-}a_i)(r_4{-}r_1)},\phi\right]\!\!\Bigg)\mathfrak{pb}_i\,.\nonumber
\end{align}
From these two contributions, mere pattern recognition suggests a `preferential' choice for the arbitrary point $\r{\alpha^*}$. In particular, if we were to take $\r{\alpha^*}$ to be $\r{r_4}$, all the terms involving $K[\phi]\times\mathfrak{pb}_i$ will cancel. As $\r{\alpha^*}$ is indeed arbitrary, this would result in a final, more compact expression:\\[-14pt]
\begin{align}\mathfrak{e}_a=&\frac{4\,i}{\sqrt{(r_3{-}r_2)(r_4{-}r_1)}}\Bigg(K[\phi]\,\,\hat{\mathfrak{db}}(\r{\alpha^*}\!\!\to \r{r_4})\hspace{-50pt}\label{simple_form_of_ls}\\
&+\sum_{i=1}^{24}\frac{y(a_i)(r_4{-}r_2)}{(r_2{-}a_i)(r_4{-}a_i)}\Pi\left[\frac{(r_4{-}a_i)(r_2{-}r_1)}{(r_2{-}a_i)(r_4{-}r_1)},\phi\right]\mathfrak{pb}_i\!\!\Bigg).\nonumber
\end{align}
(The reader may be worried about the fact that taking $\r{\alpha^*}$ to be $\r{r_4}$ sets $y(\r{\alpha^*})\to0$. As such, the first term in (\ref{simple_form_of_ls}) may appear ill-defined. However, $y(\r{\alpha^*})$ \emph{also} appears (manifestly) in the denominator in the definition of the differential form $\mathfrak{db}(\r{\alpha^*})$; as such, the evaluation---for $\hat{\mathfrak{db}}(\r\alpha)$---may be performed without taking limits (and turns out to be extremely stable, numerically). This simplified form is included in the ancillary files for this work.) 

It is worth noting that, upon any physical degeneration (at co-dimension one), the elliptic function $\mathfrak{e}_a$ in fact vanishes identically. This can be understood by noting that any such physical degeneration would correspond to an ordinary `residue' contour about the simple pole generated by the collision of the roots---($(2\pi i)$ times the \emph{residue}) about the point $\r\alpha=r_1$ or $r_4$; as there is no corresponding on-shell function to draw, the amplitude must vanish on such a contour. We have checked that it does.

The $b$-cycle elliptic leading singularity---the one encircling a branch-cut between the roots $r_1$ and $r_3$, say (and \emph{not} encircling any of the simple poles)---is easily obtained from our work above (replacing $r_2\leftrightarrow r_3$ in (\ref{first_kind_period_a}) and (\ref{third_kind_period_a})). Using these expressions for the $b$-cycle integrals of the relevant differential forms and choosing $\r{\alpha^*}$ to be $\r{r_4}$ as before, the resulting expression becomes\\[-14pt] 
\begin{align}\mathfrak{e}_b=&\frac{4}{\sqrt{(r_3{-}r_2)(r_4{-}r_1)}}\Bigg(K[1{-}\phi]\,\,\hat{\mathfrak{db}}(\r{\alpha^*}\!\!\to \r{r_4})\hspace{-50pt}\label{simple_form_of_ls_contour_b}\\
&+\sum_{i=1}^{24}\frac{y(a_i)(r_4{-}r_3)}{(r_3{-}a_i)(r_4{-}a_i)}\Pi\left[\frac{(r_4{-}a_i)(r_3{-}r_1)}{(r_3{-}a_i)(r_4{-}r_1)},1{-}\phi\right]\mathfrak{pb}_i\!\!\Bigg).\nonumber
\end{align}

One interesting aspect of these formulae is that the $b$-cycle leading singularity $\mathfrak{e}_b$ is invariant under a 5-fold cyclic rotation (in a highly non-trivial way), while the $a$-cycle integral, $\mathfrak{e}_a$, is not---reflecting the asymmetry of the contour (analogous to the non-cyclic invariance of the four-mass box leading singularities). 

Explicit, computer-usable (\textsc{Mathematica}) expressions for both elliptic leading singularities $\mathfrak{e}_a$ and $\mathfrak{e}_b$ are included as ancillary files to this work. This code makes use of tools made available in \cite{Bourjaily:2010wh,Bourjaily:2013mma,Bourjaily:2015jna}.

%%%%%%%%%%%%%%%%%%%%%%%%%%%%%%%%%%%%%%%%%%%%%%%%%%%%%%%%%%%%%%%%%%%%%%%%%%%%%%%%%%%%%%%%%%%
\vspace{-12pt}\section{Yangian Invariance of the Elliptic Leading Singularities}\vspace{-14pt}
%%%%%%%%%%%%%%%%%%%%%%%%%%%%%%%%%%%%%%%%%%%%%%%%%%%%%%%%%%%%%%%%%%%%%%%%%%%%%%%%%%%%%%%%%%%
%
Among the most interesting aspects of our results so far is that Yangian-invariance \emph{requires} the complete elliptic integrals $K[\phi]$ and $\Pi[q,\phi]$ as coefficients appearing in their definition. The easiest way to see this is to consider one of the level-one generators (see e.g.\ \cite{Drummond:2010qh})
\eq{J^A_{\phantom{A}B}\equivR\sum_{a=1}^n\,\mathcal{Z}^A_a\frac{\partial}{\partial\,\mathcal{Z}_a^B}\,,\label{Yangian_generator}}
where $\mathcal{Z}_a$ is a super-momentum-twistor, and the component $A$ is taken to be fermionic and $B$ bosonic. This operator turns out to be surprisingly powerful. For example, it tells us that \emph{any} non-trivial function of cross-ratios times a Yangian-invariant will not be Yangian-invariant. In particular, direct application of this operator demonstrates that the four-mass box coefficient as defined in \cite{Bourjaily:2013mma}, which is not simply a product of $R$-invariants but includes as part of its definition a particular function of the relevant cross-ratios, is \emph{only} Yangian-invariant with these peculiar prefactors included. 

It turns out that no combination or subset of the terms  (with constant coefficients) that appear in the two formulae for $\mathfrak{e}_{a,b}$ in (\ref{simple_form_of_ls}) and (\ref{simple_form_of_ls_contour_b}), respectively, is Yangian-invariant except for the $\mathfrak{e}_{a,b}$ themselves. We have checked this explicitly using numerical approximations for the derivatives appearing in the Yangian-generator (\ref{Yangian_generator}). 

That any integral over a compact, full-dimensional cycle in the Grassmannian should be Yangian invariant may not be surprising: indeed, it seems to be a consequence of the arguments described in \mbox{ref.\ \cite{Drummond:2010uq}}. However, the fact that these integrals, in the case of elliptic contours, \emph{require} the non-algebraic content of complete elliptic integrals is very surprising, as this is in stark contrast with the notion of taking residues (a purely algebraic operation). 

The lesson here has very obvious consequences for generalization beyond elliptic contours---which we must leave to future work.

%%%%%%%%%%%%%%%%%%%%%%%%%%%%%%%%%%%%%%%%%%%%%%%%%%%%%%%%%%%%%%%%%%%%%%%%%%%%%%%%%%%%%%%%%%%
\vspace{-16pt}\section{Conclusions and Discussion}\vspace{-14pt}
%%%%%%%%%%%%%%%%%%%%%%%%%%%%%%%%%%%%%%%%%%%%%%%%%%%%%%%%%%%%%%%%%%%%%%%%%%%%%%%%%%%%%%%%%%%
%
In this Letter, we have motivated a broader definition of `leading singularity' to be any full-dimensional compact, contour integral of a perturbative amplitude's loop integrand. This definition differs from previous one merely by factors of $(2\pi i)$ in the case of simple (logarithmic) poles, but includes also the contour integrals of elliptic curves. We have given closed formulae for elliptic-containing contour integrals for the first non-polylogarithmic structure that arises in planar sYM theory, and we have checked that they are Yangian-invariant. We conjecture that, with this broader definition, all leading singularities of sYM will be Yangian invariant, and that the set of all leading singularities will represent complete information for perturbative amplitudes in this theory. We will have more to say about this in a forthcoming work.  

\nopagebreak

%%%%%%%%%%%%%%%%%%%%%%%%%%%%%%%%%%%%%%%%%%%%%%%%%%%%%%%%%%%%%%%%%%%%%%%%%%%%%%%%%%%%%%%%%%%
\vspace{-12pt}\section{Acknowledgments}\vspace{-15pt}
%%%%%%%%%%%%%%%%%%%%%%%%%%%%%%%%%%%%%%%%%%%%%%%%%%%%%%%%%%%%%%%%%%%%%%%%%%%%%%%%%%%%%%%%%%%
The authors gratefully acknowledge fruitful conversations with Lauren Altman, Nima Arkani-Hamed, Claude Duhr, Song He, Cristian Vergu, and Anastasia Volovich. 
This work was performed in part at the Aspen Center for Physics, which is supported by National Science Foundation grant PHY-1607611, and the Harvard Center of Mathematical Sciences and Applications. 
This project has been supported by an ERC Starting Grant \mbox{(No.\ 757978)}, a grant from the Villum Fonden \mbox{(No.\ 15369)}, by a grant from the Simons Foundation (341344, LA) (JLB), and was supported in part by the US Department of Energy under contract DE-SC0010010 Task A (MS). 

\vspace{-14pt}
%\bibliographystyle{physics}
%\bibliography{amplitude_refs}
%\end{document}

\providecommand{\href}[2]{#2}\begingroup\raggedright\endgroup

\end{document}